\documentclass[reqno,12pt]{amsart}

\usepackage{amsmath}
\usepackage{amssymb}
\usepackage{hyperref}
\usepackage[mathscr]{eucal}
\usepackage{graphicx}
\newcommand{\be}{\begin{eqnarray}}
\newcommand{\ee}{\end{eqnarray}}

\begin{document}

\centerline{\bf An Instability of the Standard Model}
\centerline{\bf  Creates the Anomalous Acceleration}
\centerline{\bf  Without Dark Energy}
\vspace{.3cm}

\vspace{.3cm}

\centerline{December 8, 2014}

$$\begin{array}{ccc}
Joel\ Smoller\footnotemark[1]  &  Blake\ Temple\footnotemark[2] & Zeke\ Vogler\footnotemark[2] \nonumber \end{array}$$ 
\footnotetext[1]{Department of Mathematics, University of Michigan, Ann Arbor, MI 48109; Supported by NSF Applied Mathematics Grant Number DMS-060-3754.}

\footnotetext[2]{Department of Mathematics, University of California, Davis, Davis CA
95616; Supported by NSF Applied Mathematics Grant Number DMS-010-2493.}

\newtheorem{Theorem}{Theorem}
\newtheorem{Lemma}{Lemma}
\newtheorem{Proposition}{Proposition}
\newtheorem{Corollary}{Corollary}


\section{Abstract}
\setcounter{equation}{0}

We introduce a new asymptotic ansatz for spherical perturbations of the Standard Model of Cosmology (SM) which applies  during the $p=0$ epoch, and prove that these perturbations trigger instabilities in the SM on the scale of the supernova data.  These instabilities create a large, central region of uniform under-density which expands faster than the SM, and this central region of accelerated uniform expansion introduces into the SM {\it precisely} the same range of corrections to redshift vs luminosity as are produced by the cosmological constant in the theory of  Dark Energy.  A universal behavior is exhibited because all sufficiently small perturbations evolve to a single stable rest point.  Moreover, we  prove that these perturbations are consistent with, and the instability is triggered by, the one parameter family of self-similar waves which the authors previously proposed as possible time-asymptotic wave patterns for perturbations of the SM at the end of the  radiation epoch.   Using numerical simulations, we calculate the unique wave in the family that accounts for the same values of the Hubble constant and quadratic correction to redshift vs luminosity as in a universe with seventy percent Dark Energy, $\Omega_{\Lambda}\approx.7$.   A numerical simulation of the third order correction associated with that unique wave establishes a testable prediction that distinguishes this theory from the theory of Dark Energy.  This explanation for the anomalous acceleration, based on instabilities in the SM together with simple wave perturbations from the radiation epoch that trigger them, provides perhaps the simplest mathematical explanation for the anomalous acceleration of the galaxies that does not invoke Dark Energy.


\section{Introduction}
\setcounter{equation}{0}

In this announcement we accomplish the program set out by the first two authors in \cite{temptalk,smolte2,smolte3}, to evolve a one parameter family of GR self-similar waves which the authors identified as canonical perturbations of the Standard Model of Cosmology (SM)\footnote{Assuming the so-called Cosmological Principle, that the universe is uniform on the largest scale, the evolution of the universe on that scale is described by a Friedmann spacetime, \cite{wein}, which is  determined by the equation of state in each epoch. In this paper we let SM denote the approximation to the Standard Model of Cosmology without Dark Energy given by the critical $k=0$ Friedmann universe with equation of state $p=\frac{c^2}{3}\rho$ during the radiation epoch, and $p=0$ thereafter, (c.f. the $\Lambda CDM$ model with $\Lambda=0$, \cite{long}).} during the radiation epoch of the Big Bang, up through the  $p=0$ epoch to present time, with the purpose of investigating a possible connection with the observed anomalous acceleration of the galaxies, \cite{smolte2,long}.  Our analysis is based on the discovery of a closed ansatz for perturbations of the SM during the $p=0$ epoch of the Big Bang which triggers instabilities that create unexpectedly large regions of accelerated uniform expansion within Einstein's original theory without the cosmological constant.  We prove that these accelerated regions introduce {\it precisely} the same range of corrections to redshift vs luminosity as are produced by the cosmological constant in the theory of  Dark Energy.   A universal behavior is exhibited because all sufficiently small perturbations tend time asymptotically to a single stable rest point where the spacetime is Minkowski.  Based on this, we accomplish our initial program by  proving that these perturbations are consistent with, and the instability is triggered by, the one parameter family of self-similar waves proposed by the authors in \cite{smolte2} as possible time-asymptotic wave patterns for perturbations of the SM at the end of the  radiation epoch.  By numerical simulation we identify a unique wave in the family that accounts for the same values of the Hubble constant and quadratic correction to redshift vs luminosity as are implied by the theory of Dark Energy with $\Omega_{\Lambda}\approx.7$.   A numerical simulation of the third order correction associated with that unique wave establishes a testable prediction that distinguishes this theory from the theory of Dark Energy, (see Figure 1 below).  The result is an alternative, testable mathematical explanation for the anomalous acceleration of the galaxies that does not invoke Dark Energy, and is based upon the identification of a new instability of the SM on the scale of the supernova data, and a family of simple wave perturbations that trigger it. 

Most of the expansion of the universe before the pressure drops to $p\approx0$, is governed by the radiation epoch, a period in which the evolution is described by the equations of pure radiation.  These equations take the form of the relativistic $p$-system of shock wave theory, and for such highly nonlinear equations, one expects complicated solutions to become simpler.  A rigorous theory in one space dimension shows that solutions decay to a concatenation of {\it simple waves}, solutions along which the equations reduce to ODE's, \cite{lax,glim,smolte3}.  Based on this, together with the fact that large fluctuations from the radiation epoch (like the baryonic acoustic oscillations) are typically spherical, \cite{long}, the authors began the program in \cite{temptalk} by looking for a family of spherically symmetric solutions that perturb the SM during the radiation epoch when the equation of state $p=\frac{c^2}{3}\rho$ holds, and on which the Einstein equations reduce to ODE's.  In \cite{smolte2,smolte3}, we identified a unique family of such solutions  which we refer to as $a$-waves, parameterized by the so called {\it acceleration parameter} $a>0$, normalized so that $a=1$ is the SM.  This family of waves was first discovered from a different point of view in the profoundly interesting paper \cite{cahita} by Cahill and Taub,\footnote{Our hypotheses here are consistent with, but different from, the so-called {\it self-similarity hypothesis}, (c.f. \cite{carrco}), and a similar proposal therein to explain voids between galaxies.   As far as we know, our's is the first attempt to connect this family of waves with the anomalous acceleration.} and is the only known family of solutions which both (1) perturb Friedmann spacetimes, and (2) reduce the Einstein equations to ODEs, \cite{cahita,smolte3,carrco}.  Since when $p=0$, {\it under-densities} relative to the SM are a natural mechanism for creating anomalous {\it accelerations}, (less matter present to slow the expansion implies a larger expansion rate, \cite{long}),   we restrict to the perturbations $a<1$ which induce  under-densities relative to the SM, \cite{smolte2,smolte3}.  This requires that we abandon the Cosmological Principle on the scale of these perturbations.

Thus our starting hypothesis is that the anomalous acceleration of the galaxies is due to a local under-density relative to the SM, on the scale of the supernova data \cite{cliffela}, created by a perturbation that has decayed (locally near the center) to an $a$-wave, $a<1$, by the end of the radiation epoch.\footnote{Since time asymptotic wave patterns for the $p$-system typically involve multiple simple waves, we make no hypothesis regarding the space-time far from the center of the $a$-wave, taking the secondary waves as unknown.  This is consistent with, but modifies the self-similarity hypothesis in \cite{carrco}, and the under-density theories proposed and cited in \cite{cliffela}.}    
We prove the following:  (i)  The SM is {\it unstable} to perturbation by $a$-waves;  (ii) A small under-density created by an $a$-wave at the end of radiation, triggers the formation of a large region of accelerated expansion which extends further and further outward from the center, becoming more flat and more uniform, as time evolves; (iii)  Neglecting errors in the density of fourth order in fractional distance to the Hubble Length, this extended region moving outward from the center evolves according to an autonomous system of two ODE's, and is described by a solution trajectory that starts near a classic {\it unstable saddle rest point} corresponding to the SM at the end of radiation, and evolves to a nearby {\it stable} rest point where the metric is Minkowski.   During this evolution,  the quadratic correction to redshift vs luminosity (as measured near the center) assumes {\it precisely the same range of values} as Dark Energy theory.  That is, letting
\begin{eqnarray}\label{redvslum}
H\,d_{\ell}={\rm z}+Q{\rm z}^2+C{\rm z}^3+O({\rm z}^4)
\end{eqnarray}
denote  the relation between redshift factor ${\rm z}$ and luminosity distance $d_{\ell}$ at a given value of the Hubble constant $H$ as measured at the center\footnote{For the FRW spacetime, $Q$ is determined by the value of the so-called {\it deceleration parameter} $q$, and $C$ is determined by the {\it  jerk} $j$, c.f., \cite{long}.},  the value of the quadratic correction $Q$ increases from  the SM value $Q=.25$ at the end of radiation, to the value $Q=.5$ as $t\rightarrow\infty.$   This is precisely the same range of values $Q$ takes on in Dark Energy theory as the fraction $\Omega_{\Lambda}$ of Dark Energy to classical energy increases from its value of $\Omega_{\Lambda}\approx0$ at the end of radiation, to $\Omega_{\Lambda}=1$ as $t\rightarrow\infty$.  This holds for any $a<1$ near $a=1$, and for any value of the cosmological constant $\Lambda>0$, assuming only that $a$ and $\Lambda$ both induce a negligibly small correction to the SM value $Q=.25$ at the end of radiation.\footnote{We qualify with this latter assumption only because, in Dark Energy theory, the value of $\Omega_{\Lambda}$ is small but not exactly equal to zero at the end of radiation; and in the wave theory, the value of $Q$ jumps down slightly below $Q=.25$ at the end of radiation before it increases to $Q=.5$ from that value as $t\rightarrow\infty$.}

These results are recorded in the following theorem.  Here we let present time in a given model denote the time at which the Hubble constant $H$ (as defined in (\ref{redvslum}))  reaches its present measured value $H=H_0$, this time being different in different models.  
\begin{Theorem}\label{Thmintro}
Let $t=t_{0}$ denote present time in the wave model and $t=t_{DE}$ present time in the Dark Energy\,\footnote{By the Dark Energy model we refer to the critical $k=0$ Friedmann universe with cosmological constant, taking the present value $\Omega_{\Lambda}=.7$ as the best fit to the supernova data among the two parameters $(k,\Lambda)$,\cite{perl,perl2}.} model.  Then there exists a unique value of the acceleration parameter $\underbar{\it a}=0.99999959\approx 1-4.3\times10^{-7}$  corresponding to an under-density relative to the SM at the end of radiation, such that the subsequent $p=0$ evolution starting from this initial data evolves to time $t=t_{0}$ with $H=H_0$ and $Q=.425$, in agreement with the values of $H$ and $Q$ at $t=t_{DE}$ in the Dark Energy model.    The cubic correction at $t=t_{0}$ in the wave theory is then  $C=0.359$, while Dark Energy theory gives $C=-0.180$ at $t=t_{DE}$.  The times are related by $t_0\approx 1.45\,t_{DE}$.  
\end{Theorem}

We emphasize that $t_0$, $Q$ and $C$ in the wave model,  are determined by $a$ alone.  Indeed,  the initial data at the end of radiation, which determines the $p=0$ evolution, depends, at the start, on two parameters: the acceleration parameter $a$ of the self-similar waves, and the initial temperature $T_*$ at which the pressure is assumed to drop to zero.  But our numerics show that the dependence on the starting temperature is negligible for $T_*$ in the range $3000^oK\leq T_*\leq 9000^oK,$ (the range assumed in cosmology, \cite{long}).  Thus for the temperatures appropriate for Cosmology, $t_0$, $Q$ and $C$ are determined by $a$ alone. 

A measure of the {\it severity} of the instability created by the $a=\underbar{\it a}$ perturbation of the SM, is quantified by  the numerical simulation.   For example, comparing the initial density $\rho_{wave}$ for $a=\underbar{\it a}$ at the center of the wave, to the corresponding initial density $\rho_{SM}$ in the SM  at the end of radiation $t=t_*$, gives $\frac{\rho_{wave}}{\rho_{SM}}\approx 1-(7.45)\times 10^{-6}\approx1$.   During the $p=0$ evolution, this ratio evolves to a {\it seven-fold} under-density in the wave model relative to the SM by present time, i.e., $\frac{\rho_{wave}}{\rho_{SM}}=0.146$ at $t=t_0$.  
    
Note that in principle adding acceleration to a model should increase the expansion rate $H$ and consequently the age of the universe, because it then takes longer for the Hubble constant $H$ to decrease to its present small value $H_0$.    Incorporating Dark Energy taking $\Omega_{\Lambda}=.7$ increases the age $t_{SM}=.96\times 10^{10}yr$ of SM by about $45\%$ to $t_{DE}= 1.39\times10^{10}yr$, and the wave theory increases it by another $45\%$ to $t_0= 2.03\times10^{10}yr$.  

Our wave theory is based on the self-similarity variable $\xi=r/ct<1$, which measures the fractional distance from the center $\xi=0$ to the Hubble length at time $t$.\footnote{Here we let $t$ and $r$ denote time and radial coordinates in Standard Schwarzschild Coodinates (SSC) in which $r$ is arclength distance at fixed $t$, \cite{wein,smolte,smolte3}}   We show below (c.f. Section \ref{S2.3}), that if we neglect errors $O(\xi^4),$ and then further neglect small errors between the wave metric and the Minkowski metric (which tend to zero, at that order, with approach to the stable rest point, c.f. (iii) above), and also neglect errors due to relativistic corrections in the velocities of the fluid relative to the center (where the velocity is zero), the resulting spacetime is,  like a Friedmann spacetime,  {\it independent of the choice of center}. 
Thus the central region of approximate uniform density at present time $t=t_0$ in the wave model extends out from the center $r=0$ at $t=0$ in SSC, to radial values $r$ small enough so that the fractional distance to the Hubble length $\xi=r/ct_0$ satisfies $\xi^4<<1$.  We conclude that since the age of the universe $t_0$ in the wave model is about twice as old as in the SM {\it without DE}, it follows that the central region of uniform acceleration in the wave model would be about {\it twice as large} as the SM age $t_{SM}$ would predict.

  The cubic correction $C$ to redshift vs luminosity is a {\it verifiable} prediction of the wave theory that distinguishes it from Dark Energy theory.  In particular, $C>0$ in the wave model and $C<0$ in the Dark Energy model implies that the cubic correction {\it increases} the right hand side of (\ref{redvslum}), (i.e., increases the discrepancy between the observed redshifts and the predictions of the SM) far from the center in the wave theory, while it {\it decreases} the right hand side of (\ref{redvslum}) far from the center in  the Dark Energy theory.  Now the anomalous acceleration was originally derived  from a collection of data points, and the $\Omega_{\Lambda}\approx.7$ critical FRW spacetime is obtained as the best fit to Friedmann spacetimes among the parameters $(k,\Lambda)$.  We understand that the current data  is sufficient to provide a value for $Q$, but not $C$, \cite{hugh}.   It is not clear to the authors whether or not there are indications in the data that could distinguish $C<0$ from $C>0$.

\section{Presentation of Results}\label{S2.0}   

We summarize the sections of our forthcoming paper which  brings our identified one parameter family of GR $a$-waves up through the $p=0$ epoch of the Big Bang to present time. We quantify the quadratic corrections $Q$ implied in (\ref{redvslum}) by these perturbations to SM near the center, and compare the results with Dark Energy theory.  
 
We begin by recalling that $a$-waves form a $1$-parameter family of spherically symmetric solutions of the Einstein equations $G=\kappa T$ that depend only on the self-similarity variable $\xi=r/t$, and exist when $p=\frac{c^2}{3}\rho$.  They reduce to the critical SM Friedmann spacetime for pure radiation when $a=1$.   In contrast, when $p=0$, only the SM $p=0$, $k=0$ Friedmann spacetime can be expressed in this self-similar form.  Our expansion of the time independent self-similar waves during the radiation epoch in powers of $\xi$ calculated in \cite{smolte3} has led us to the discovery of a new {\it time dependent} asymptotic ansatz for corrections to the standard model, that depend on $(t,\xi)$, and close at order $\xi^4$ under the $p=0$ evolution.  This ansatz is sufficiently general to incorporate initial data from the self-similar waves at the end of radiation, and hence the evolution of these waves into time dependent solutions during the $p=0$ epoch.  In this paper we deduce the evolution of the corrections induced by $a$-waves at the end of radiation from the phase portrait of these asymptotic equations. In fact, our main result, that an instability in the SM can create the anomalous acceleration of the galaxies without Dark Energy, applies not just to perturbations by $a$-waves, but to any perturbation  consistent with our asymptotic ansatz at the end of radiation, so long as the perturbation lies within the domain of attraction of the stable rest point to which the perturbation $a=\underline{\it a}$ evolves.  

In Sections \ref{S2.1}-\ref{S2.3} we derive an alternative formulation of the $p=0$ Einstein equations in spherical symmetry, introduce our new asymptotic ansatz for corrections to the SM, use the exact equations to derive equations for the corrections, and use these to characterize the instability.  In Section \ref{S2.4} we derive the correct redshift vs luminosity relation for the SM including the corrections.  In Section \ref{S2.5} we introduce a gauge transformation that converts the $a$-waves at the end of radiation into initial data that is consistent with our ansatz.  In Section \ref{S2.6} we present our numerics that identifies the unique $a$-wave $a=\underline{\it a}$ in the family that meets the conditions $H=H_0$ and $Q=.425$ at $t=t_0$,  and explain our predicted cubic correction $C=0.359$.   In  Section \ref{S2.7} we discuss the uniform space-time created at the center of the perturbation. Concluding remarks are given in Section \ref{S4}.  Details are omitted in this announcement.  We use the convention $c=1$ when convenient.   
   
\subsection{The $p=0$ Einstein Equations in Coordinates Aligned with the Physics}\label{S2.1}

In this section we introduce  a new formulation of the $p=0$ Einstein equations that describe outwardly expanding spherically symmetric solutions.   We do not employ co-moving coordinates, \cite{cliffela}, but rather use $\xi$ as a spacelike variable because it is better aligned with the physics.  That is, our derivation starts with metrics in Standard Schwarzschild Coordinates (SSC), where the metric takes the canonical form,
\begin{eqnarray}
ds^2=-B(t,r)dt^2+\frac{1}{A(t,r)}dr^2+r^2d\Omega^2,
\end{eqnarray}
 but our subsequent analysis is done in $(t,\xi)$ coordinates, where $\xi=r/t$.   Our starting point is the observation that the SSC metric form is invariant under transformations of $t$, and there exists a time coordinate in which SM is self-similar in the sense that the metric components $A,B$, the velocity $v$ and $\rho r^2$ are functions of $\xi$ alone.   This self-similar form exists, but is different for $p=\frac{c^2}{3}\rho$ and  $p=0$, \cite{carrco,smoltevo}. Taking $p=0$, letting $v$ denote the SSC velocity and $\rho$ the co-moving energy density, and eliminating all unknowns in terms of $v$ and the Minkowski energy density $T^{00}_{M}=\frac{\rho}{1-\left(\frac{v}{c}\right)^2}$, (c.f. \cite{groate}), the locally inertial formulation of the Einstein equations $G=\kappa T$ introduced in \cite{groate} reduce to 

\begin{eqnarray}\nonumber
&\left(\kappa T^{00}_{M}r^2\right)_t+\left\{\sqrt{AB}\frac{v}{r}\left(\kappa T^{00}_{M}r^2\right)\right\}_r=-2\sqrt{AB}\frac{v}{r}\left(\kappa T^{00}_{M}r^2\right),\ \ \ \ \ \ \\\nonumber
&\ \ \ \left(\frac{v}{r}\right)_t+r\sqrt{AB}\left(\frac{v}{r}\right)\left(\frac{v}{r}\right)_r=-\sqrt{AB}
\left\{\left(\frac{v}{r}\right)^2+\frac{1-A}{2Ar^2}\left(1-r^2\left(\frac{v}{r}\right)^2\right)\right\},\\\nonumber
&r\frac{A'}{A}=\left(\frac{1}{A}-1\right)-\frac{1}{A}\kappa T^{00}_{M}r^2,\ \ \ \ \ \ \\\nonumber
&r\frac{B'}{B}=\left(\frac{1}{A}-1\right)+\frac{1}{A}\left(\frac{v}{c}\right)^2\kappa T^{00}_{M}r^2,
\end{eqnarray}
where prime denotes $d/dr.$   Note that the $1/r$ singularity is present in the equations  because incoming waves can amplify without bound.  We resolve this for outgoing expansions by assuming $w=v/\xi$ is positive and finite at $r=\xi=0$.   Making the substitution $D=\sqrt{AB}$, taking $z=\kappa T^{00}_{M}r^2$ as the dimensionless density, $w=\frac{v}{\xi}$ as the dimensionless velocity with $\xi= r/t$  and rewriting the equations in terms of $(t,\xi)$, we obtain
\begin{eqnarray}\nonumber
&tz_t+\xi\left\{(-1+Dw)z\right\}_\xi=-Dwz,\ \ \ \ \ \ \ \ \ \ \ \ \ \ \ \ \ \ \ \ \ \ \\\nonumber
&\ \ \ tw_t+\xi\left(-1+Dw\right)w_\xi=w-D\left\{w^2+\frac{1-\xi^2w^2}{2A}\left[\frac{1-A}{\xi^2}\right]\right\}
\ \ \ \ \\\nonumber
&\xi A_{\xi}=\left(A-1\right)-z\ \ \ \ \ \ \ \ \ \ \ \ \ \ \ \ \ \ \ \ \ \ \ \ \ \ \\\nonumber
&\frac{\xi D_{\xi}}{D}=\left(A-1\right)-\frac{\left(1-\xi^2w^2\right)}{2}z.\ \ \ \ \ \ \ \ \ \ \ \ \ \ \ 
\end{eqnarray}
That is, since the sound speed is zero when $p=0$,  $w(t,0)>0$ restricts us to expanding solutions in which all information from the fluid propagates outward from the center.  (Cusp singularities in the velocity at $r=0$ in SSC are regularized in co-moving coordinates, \cite{smolte3}.)

\subsection{A New Ansatz for Corrections to SM}\label{S2.2}

We introduce the following ansatz for corrections to SM near $\xi=0$ that involves only even powers of $\xi$, where we can interpret $\xi=r/ct\equiv r/t$ as a measure of the fractional distance to the Hubble length, \cite{smolte3,smoltevo}:  
\begin{eqnarray}
&z(t,\xi)=z_{SM}(\xi)+\Delta z(t,\xi)\ \ \ \ \ \ \ \ \ \Delta z=z_2(t)\xi^2+z_4(t)\xi^4\ \ \ \ \ \label{ansatz1}\\
&w(t,\xi)=w_{SM}(\xi)+\Delta w(t,\xi) \ \ \ \ \ \ \ \Delta w=w_0(t)+w_2(t)\xi^2\ \ \ \ \ \ \ \label{ansatz2}\\
&A(t,\xi)=A_{SM}(\xi)+\Delta A(t,\xi)\ \ \ \ \ \ \ \Delta A=A_2(t)\xi^2+A_4(t)\xi^4\ \ \ \ \label{ansatz3}\\
&D(t,\xi)=D_{SM}(\xi)+\Delta D(t,\xi)\ \ \ \ \ \ \Delta D=D_2(t)\xi^2\ \ \ \ \ \ \ \ \ \ \ \ \ \ \ \ \ \label{ansatz4}
\end{eqnarray}
where $z_{SM}, w_{SM}, A_{SM}, D_{SM}$ are the expressions for the unique self-similar representation of the SM when $p=0$, given by, \cite{smoltevo},
\begin{eqnarray}\label{ansatzzw}
&z_{SM}(\xi)=\frac{4}{3}\xi^2+\frac{40}{27}\xi^4+O(\xi^6),\ \ w_{SM}(\xi)=\frac{2}{3}+\frac{2}{9}\xi^2+O(\xi^4),\\ 
&A_{SM}(\xi)=1-\frac{4}{9}\xi^2-\frac{8}{27}\xi^4+O(\xi^6),\ \ \ 
D_{SM}(\xi)=1-\frac{1}{9}\xi^2+O(\xi^4).\ \ \label{ansatzAD}
\end{eqnarray}
This gives
\begin{eqnarray}
z(t,\xi)&=&\left(\frac{4}{3}+z_2(t)\right)\xi^2+\left\{\frac{40}{27}+z_4(t)\right\}\xi^4+O(\xi^6),\nonumber\label{ansatzz}\\
w(t,\xi)&=&\left(\frac{2}{3}+w_0(t)\right)+\left\{\frac{2}{9}+w_2(t)\right\}\xi^2+O(\xi^4).\nonumber\label{ansatzw}
\end{eqnarray}
We prove the equations close within this ansatz, at order $\xi^4$ in $z$ and order $\xi^2$ in $w$, with errors $O(\xi^6)$ in $z$ and $O(\xi^4)$ in $w$.    Moreover, the importance of this ansatz is that corrections satisfying the ansatz induce an instability in the SM by creating a uniform spacetime of density $\rho(t)$, constant at each fixed $t$, out to errors of order $O(\xi^4)$.  That is, since the ansatz,  
\begin{eqnarray}\label{uniformdensity}
z(\xi,t)=\kappa\rho(t,\xi)r^2+O(\xi^4)=\left(\frac{4}{3}+z_2(t)\right)\xi^2+O(\xi^4),
\end{eqnarray}
neglecting the $O(\xi^4)$ error gives $\kappa\rho=(4/3+z_2(t))/t^2$, a function of time alone.   For the SM, $z_2\equiv0$ and this gives $\kappa\rho(t)=\left(4/3\right)t^{-2}$, which is the exact evolution of the density for the SM Friedmann spacetime with $p=0$ in co-moving coordinates, \cite{smolte}.  For the evolution of our specific under-densities in the wave theory, we show $z_2(t)\rightarrow-4/3$ as the solution tends to the stable rest point, implying that the instability creates an accelerated drop in the density in a large uniform spacetime expanding outward from the center.   (See Section \ref{S2.7} below.)

\subsection{Asymptotic equations for Corrections to SM}\label{S2.3}

Substituting the ansatz (\ref{ansatz1})-(\ref{ansatz4}) for the corrections into the Einstein equations $G=\kappa T$, and neglecting terms $O(\xi^4)$ in $w$ and $O(\xi^6)$ in $z$, we obtain the following closed system of ODE's for the corrections $z_2(\tau)),$ $z_4(\tau),$ $w_0(\tau),$ $w_2(\tau)$, where $\tau=\ln{t}$, $0<\tau\leq 11$.  (Introducing $\tau$ renders the equations autonomous, and solves the long time simulation problem.)  Letting prime denote $d/d\tau$, the equations for the corrections reduce to the autonomous system 
\begin{eqnarray}\label{z2finalsumeqn112}
z'_2&=&-3w_0\left(\frac{4}{3}+z_2\right),\\\label{w0finalsumeqn112}
w'_0&=&-\frac{1}{6}z_2-\frac{1}{3}w_0-w_0^2,\\\label{z4finalsumeqn112}
z'_4&=&5\left\{\frac{2}{27}z_2+\frac{4}{3}w_2-\frac{1}{18}z_2^2+z_2w_2\right\}\\\nonumber
&&\ \ \ \ \ \ \ \ \ \ \ \ \ \ \ \ \ \ \ \ \ +5w_0\left\{\frac{4}{3}-\frac{2}{9}z_2+z_4-\frac{1}{12}z_2^2\right\},\\\label{w2finalsumeqn112}
w'_2&=&-\frac{1}{10}z_4-\frac{4}{9}w_0+\frac{1}{3}w_2-\frac{1}{24}z_2^2+\frac{1}{3}z_2w_0\\\nonumber
&&\ \ \ \ \ \ \ \ \ \ \ \ \ \ \ \ \ \ \ \ \ \ \ \ \ +\frac{1}{3}w_0^2-4w_0w_2+\frac{1}{4}w_0^2z_2.
\end{eqnarray}
We prove that for the equations to close within the ansatz (\ref{ansatz1})-(\ref{ansatz4}), it is necessary and sufficient to assume the initial data satisfies the gauge conditions
\begin{eqnarray}\label{next4}
A_2=-\frac{1}{3}z_2,\ \ A_4=-\frac{1}{5}z_4,\ \ 
D_2=-\frac{1}{12}z_2. 
\end{eqnarray}
We prove that if these constraints hold initially, then they are maintained by the equations for all time. Conditions (\ref{next4}) are not invariant under time transformations, even though the SSC metric form is invariant under arbitrary time transformations, so we can interpret (\ref{next4}), and hence the ansatz (\ref{ansatz1})-(\ref{ansatz4}), as fixing the time coordinate gauge of our SSC metric.  This gauge agrees with FRW co-moving time up to errors of order $O(\xi^2)$.

The autonomous $4\times4$ system (\ref{z2finalsumeqn112})-(\ref{w2finalsumeqn112}) contains within it the closed, autonomous, $2\times2$ sub-system (\ref{z2finalsumeqn112}), (\ref{w0finalsumeqn112}). This sub-system describes the evolution of the corrections $(z_2,w_0)$, which we show in Section \ref{S2.4} determines the quadratic correction $Qz^2$ in (\ref{redvslum}).  Thus the sub-system  (\ref{z2finalsumeqn112}), (\ref{w0finalsumeqn112}) gives the corrections to SM at the order of the observed anomalous acceleration, accurate within the central region where errors $O(\xi^4)$ in $z$ and orders $O(\xi^3)$ in $v=w/\xi$ can be neglected.   The phase portrait for sub-system (\ref{z2finalsumeqn112}), (\ref{w0finalsumeqn112}) exhibits an unstable saddle rest point at $(z_2,w_0)=(0,0)$ corresponding to the SM, and a stable rest point at $(z_2,w_0)=(-4/3,1/3)$. These are the rest points referred to in the introduction.  From the phase portrait, (see Figure 1), we see that perturbations of SM corresponding to small under-densities will evolve away from the SM near the unstable manifold of $(0,0)$, and toward the stable rest point.  By (\ref{ansatzzw}) and (\ref{ansatzAD}), $A_2=4/9,D_2=1/9$ at $(z_2,w_0)=(-4/3,1/3)$, so by (\ref{ansatzAD}) the metric components $A$ and $B$ are equal to $1+O(\xi^4)$, implying the metric at the stable rest point $(-4/3,1/3)$ is Minkowski up to $O(\xi^4)$. Thus during evolution toward the stable rest point, the metric tends to flat Minkowski spacetime with $O(\xi^4)$ errors.   

\subsection{Redshift vs Luminosity Relations for the Ansatz}\label{S2.4}

In this section we obtain formulas for $Q$ and $C$ in (\ref{redvslum}) as a function of the corrections $z_2,w_0,z_4,w_2$ to the SM, we compare this to the values of $Q$ and $C$ as a function of $\Omega_{\Lambda}$ in DE theory, and we show that remarkably, $Q$ passes through the same range of values in both theories.   

Recall that $Q$ and $C$ are the quadratic  and cubic corrections to redshift vs luminosity as measured by an observer at the center of the spherically symmetric perturbation of the SM determined by these corrections.\footnote{The uniformity of the center out to errors $O(\xi^4)$ implies that these should be good approximations for observers somewhat off-center with the coordinate system of symmetry for the waves.}   The calculation requires taking account of all of the terms that affect the redshift vs luminosity relation when the spacetime is not uniform, and the coordinates are not co-moving.  

The redshift vs luminosity relation for the $k=0$, $p=\sigma\rho$, FRW spacetime, at any time during the evolution,  is given by,  
\begin{eqnarray}\label{sm0}
Hd_{\ell}=\frac{2}{1+3\sigma}\left\{(1+z)-\left(1+z\right)^{\frac{1-3\sigma}{2}}\right\},
\end{eqnarray}
where only $H$ evolves in time, \cite{gronhe}.   For pure radiation $\sigma=1/3$, which gives $H d_{\ell}=z$, and when $p=\sigma=0$, we get, (c.f. \cite{smolte3}),
\begin{eqnarray}\label{sm00}
Hd_{\ell}=z+\frac{1}{4}z^2-\frac{1}{8}z^3+O(z^4).
\end{eqnarray}
The redshift vs luminosity relation in the case of Dark Energy theory, assuming a critical Friedmann space-time with the fraction of Dark Energy $\Omega_{\Lambda}$, is
\begin{eqnarray}\label{DE4}
Hd_{\ell}=(1+z)\int_0^z\frac{dy}{\sqrt{{\mathcal E}(y)}},
\end{eqnarray}  
where
\begin{eqnarray}\label{DE5}
{\mathcal E}(z)=\Omega_{\Lambda}(1+z)^2+\Omega_M(1+z)^3,
\end{eqnarray}  
and  $\Omega_M=1-\Omega_{\Lambda}$, the fraction of the energy density due to matter, (c.f. (11.129), (11.124) of \cite{gronhe}).  
Taylor expanding gives
\small
\begin{eqnarray}\label{DE12}
Hd_{\ell}=z+\frac{1}{2}\left(-\frac{\Omega_M}{2}+1\right)z^2+\frac{1}{6}\left(-1-\frac{\Omega_M}{2}+\frac{3\Omega_M^2}{4}\right)z^3+O(z^4),
\end{eqnarray}
\normalsize 
where $\Omega_M$ evolves in time, ranging from $\Omega_M=1$ (valid with small errors at the end of radiation) to $\Omega_M=0$ (the limit as $t\rightarrow\infty$).  From (\ref{DE12}) we see that in Dark Energy theory, the quadratic term $Q$ increases exactly through the range
\begin{eqnarray}\label{range} .25\leq.Q\leq 5,
\end{eqnarray}
and the cubic term decreases from $-1/8$ to $-1/6$, during the evolution from the end of radiation to $t\to\infty$, thereby verifying the claim in  Theorem \ref{Thmintro}.
In the case $\Omega_M=.3$,  $\Omega_{\Lambda}=.7$,    
 representing present time $t=t_{DE}$ in Dark Energy theory, this gives
the exact expression,
\begin{eqnarray}\label{DE14}
H_0d_{\ell}={\rm z}+\frac{17}{40}\,{\rm z}^2-\frac{433}{2400}\,{\rm z}^3+O({\rm z}^4),
\end{eqnarray} 
verifying that $Q=.425$ and $C=-.1804$, as recorded in Theorem \ref{Thmintro}.

In the case of a general non-uniform spacetime in SSC, the formula for redshift vs luminosity as measured by an observer at the center is given by, (see \cite{gronhe}),
\begin{eqnarray}\label{r11thm}
d_{\ell}=(1+{\rm z})^2r_e=t_0(1+{\rm z})^2\xi_e\left(\frac{t_e}{t_0}\right),
\end{eqnarray}
where $(t_e,r_e)$ are the SSC coordinates of the emitter, and $(0,t_0)$ are the coordinates of the observer.   A calculation based on using the metric corrections to obtain $\xi_e$ and $t_e/t_0$ as functions of ${\rm z}$, and substituting this into (\ref{r11thm}), gives the following formula for the quadratic correction $Q=Q(z_2,w_0)$ and cubic correction $C=C(z_2,w_0,z_4,w_2)$
to redshift vs luminosity in terms of arbitrary corrections $w_0,w_2,z_2,z_4$ to SM.  We record the formulas in the following theorem:

\begin{Theorem}
Assume a GR spacetime in the form of our ansatz (\ref{ansatz1})-(\ref{ansatz4}), with arbitrary given corrections $w_0(t),w_2(t),z_2(t),z_4(t)$ to SM.  Then the quadratic and cubic corrections $Q$ and $C$ to redshift vs luminosity in (\ref{redvslum}), as measured by an observer at the center $\xi=r=0$ at time $t$, is given explicitly by 
\begin{eqnarray}\label{r10.4thm}
 Hd_{\ell}={\rm z}\left\{1+\left[\frac{1}{4}+E_2\right]{\rm z}+\left[-\frac{1}{8}+E_3\right]{\rm z}^2\right\}+O({\rm z}^4),
\end{eqnarray}
where 
\begin{eqnarray}\nonumber
H=\left(\frac{2}{3}+w_0(t)\right)\frac{1}{t},
\end{eqnarray}
so that
\begin{eqnarray}
Q(z_2,w_0)=\frac{1}{4}+E_2,\ \ \ \ \ C(w_0,w_2,z_2,z_4)=-\frac{1}{8}+E_3,
\end{eqnarray}
where $E_2=E_2(z_2,w_0),$ $E_3=E_3(z_2,w_0,z_4,w_2)$ are the corrections to the $p=0$ standard model values in (\ref{sm00}).  The function $E_2$ is given explicitly by 
\begin{eqnarray}\label{quadcorrect}
E_2=\frac{24w_0+45w_0^2+3z_2}{4(2+3w_0)^2}.
\end{eqnarray}
The function $E_3$ is defined by the following chain of variables:
 \begin{eqnarray}
E_3=2I_2+I_3,
\end{eqnarray}
\begin{eqnarray}\nonumber
 I_{2,3}=J_2+\frac{9w_0}{2(2+3w_0)},\ J_3+3\left[-1+\left(\frac{8-8J_2+3w_0-12J_2w_0}{2(2+3w_0)^2}\right)\right],
\end{eqnarray}
\begin{eqnarray}\nonumber
 &&J_{2}=\frac{1}{4}\left\{1-\frac{1+9K_2}{\left(1+\frac{3}{2}w_0\right)^{2}}\right\},\nonumber\\\nonumber
 && J_{3}=\frac{5}{8}\left\{1-\frac{1-\frac{18}{5}K_2-\frac{81}{5}K_2^2+\frac{9}{5}w_0+\frac{27}{5}K_3+\frac{81}{10}Q_3w_0}{\left(1+\frac{3}{2}w_0\right)^{4}}\right\},\ \ \ \ \ \ \ \ 
\end{eqnarray}
\normalsize
\begin{eqnarray}\nonumber
 K_{2,3}=\frac{2}{3}w_0+\frac{1}{2}w_0^2-\frac{1}{12}z_2,\ \frac{2}{9}w_0+w_0^2+\frac{1}{2}w_0^3+w_2-\frac{1}{18}z_2-\frac{1}{3}z_2w_0.
\end{eqnarray}

\end{Theorem}
From (\ref{quadcorrect}) one sees that $Q$ depends only on $(z_2,w_0)$, $Q(0,0)=.25$, (the exact value for the SM), $Q(-4/3,1/3)=.5$, (the exact value for the stable rest point), and from this it follows that $Q$ increases through precisely the same range (\ref{range}) of DE, from $Q\approx.25$ to $Q=.5$, along the orbit of (\ref{z2finalsumeqn112}), (\ref{w0finalsumeqn112}) that takes the unstable SM rest point $(z_2,w_0)=(0,0)$ to the stable rest point $(z_2,w_0)=(-4/3,1/3)$, (c.f. Figure 1).

\subsection{Initial Data from the Radiation Epoch}\label{S2.5}

 In this section we compute the initial data for the $p=0$ evolution from the restriction of the one parameter family of self-similar $a$-waves to a constant temperature surface $T=T_*$ at the end of radiation, and convert this to initial data on a constant time surface $t=t_{*}$, these two surfaces being different when $a\neq1$.\footnote{In \cite{smolte3} the authors derived a system of ODE's on which the SSC equations reduce to ODE's in the variable $\xi$ when $p=\frac{c^2}{3}\rho$, and extracted from this the one parameter family of $a$-waves.  In this section we use the expansions of $a$-waves into powers of $\xi$ computed in \cite{smolte3}.  Interestingly, there are no self-similar perturbations of the SM corresponding to $a$-waves when $p=0$, c.f. \cite{smoltevo,carrco}. } 
We then must define a gauge transformation that converts the resulting initial data to equivalent initial data that meets the gauge conditions (\ref{next4}).  (Recall that condition (\ref{next4}) fixes a time coordinate, or gauge, for the underlying SSC metric associated with our ansatz, and the initial data for the $a$-waves is given in a different gauge because time since the big bang depends on the parameter $a$, as well as on the pressure, so it changes when $p$ drops to zero.)   
The equation of state of pure radiation is derived from the the Stefan-Boltzmann Law, which relates the initial density $\rho_*$ to the initial temperature $T_*$ in degrees Kelvin by
\begin{eqnarray}
\label{SBLaw10}
\rho_*=\frac{a_sc}{4} T_*^4,
\end{eqnarray}
where $a_s$ is the Stefan-Boltzmann constant, \cite{peac}).   
According to current theories in cosmology,  (see e.g. \cite{peac}), the pressure drops precipitously to zero at a temperature $T=T_*$ somewhere between $3000^oK\leq T_*\leq 9000^oK$,  corresponding to starting times $t_*$ roughly in the range $10,000yr\leq t_*\leq 30,000yr$ after the Big Bang. We make the assumption that the pressure drops discontinuously to zero at some temperature $T_*$ within this range.  That our resulting simulations are numerically independent of starting temperature, (c.f. Section \ref{S2.6}), justifies the validity of this assumption.  Using this assumption, we can take the values of the $a$-waves on the surface $T=T_*$ as the initial data for the subsequent $p=0$ evolution.
Using the equations we convert this to initial data on a constant time surface $\bar{t}=\bar{t}_*$, where $\bar{t}$ is the time coordinate used in the self-similar expression of the $a$-waves which assumes $p=\frac{c^2}{3}\rho$.  Our first theorem proves that there is a gauge transformation $\bar{t}\rightarrow t$ which converts the initial data for $a$-waves at the end of radiation at $\bar{t}=\bar{t}_*$, to initial data that both meets the assumptions of our ansatz (\ref{ansatz1})-(\ref{ansatz4}), as well as the gauge conditions (\ref{next4}).   

\begin{Theorem}\label{Thmshorter}  Let $\bar{t}$ be the time coordinate for the self-similar waves during the radiation epoch, and define the transformation $\bar{t}\rightarrow t$ by
 \begin{eqnarray}\label{timetilde}
 t=\bar{t}+\frac{1}{2} q(\bar{t}-\bar{t}_*)^2- t_B,
\end{eqnarray}
where $ q$ and $ t_B$ are given by 
\begin{eqnarray}\label{thatsubB}\label{thatBfinal}
 t_B=\bar{t}_*(1-\frac{1}{5}\left(\frac{1+a^2}{1.3-a^2}\right),
\end{eqnarray}
\begin{eqnarray}\label{qhat0}
 q=\frac{a^2}{2(1+a^2)}.
 \end{eqnarray}
Then upon performing the gauge transformation (\ref{timetilde}), the initial data from the $a$-waves at the end of radiation $\bar{t}=\bar{t_*}$, meets the conditions for the ansatz (\ref{ansatz1})-(\ref{ansatz4}), as well as the gauge conditions (\ref{next4}).   
\end{Theorem}

Our conclusions are summarized in the following theorem:
 \begin{Theorem}
The initial data for the $p=0$ evolution determined by the self-similar $a$-wave on a constant time surface $t=t_{*}$ with temperature $T=T_*$ at $r=0$, is given as a function of the acceleration parameter $a$ and the temperature $T_*$, by
\begin{eqnarray}
&z_2( t_{*})=\hat{z}_2,\ \ \ \ \ \ \ 
z_4( t_{*})=\hat{z}_4+3\hat{w}_0\left(\frac{4}{3}+\hat{z}_2\right)\gamma,\ \ \ \ \ \ \ \nonumber\\
&w_0( t_{*})=\hat{w}_0,\ \ \ \ \ \ 
w_2( t_{*})=\hat{w}_2+\left(\frac{1}{6}\hat{z}_2+\frac{1}{3}\hat{w}_0+\hat{w}_0^2\right)\gamma,\nonumber
\end{eqnarray}
where $\hat{z}_2,\hat{z}_4,\hat{w}_0,\hat{w}_2$ and $\gamma$ are functions of acceleration parameter $a$  given by
\begin{eqnarray}\nonumber
&\hat{z}_2=\frac{3a^2\alpha^2}{4}-\frac{4}{3},\ \ \ \ \ \ \ 
\hat{z}_4=\frac{15a^2(\frac{3}{2}-a^2)\alpha^4}{16}-\frac{40}{27},\ \ \ \ 
\\\nonumber
&\hat{w}_0=\frac{\alpha}{2}-\frac{2}{3},\ \ \ \ \ \ \ \ \ 
\hat{w}_2=\frac{\alpha^3}{16}\left(9.5-8a^2\right)-\frac{2}{9},\ \
\end{eqnarray} 
where
\begin{eqnarray}
\label{gammatildenotilde}
\gamma=\frac{(1+a^2)\alpha}{8},\ \ \ \ \ \ \ \alpha=\frac{(1+a^2)}{5(1.3-a^2)}.
\end{eqnarray}
The time $t_{*}$ is then given in terms of the initial temperature $T_*$ by
\begin{eqnarray}
\label{tstartildelasttime}
 t_{*}=\frac{a\alpha}{2}\sqrt{\frac{3}{\kappa\rho_*}},\ \ \ \ \rho_*=\frac{a_s}{4c} T_*^4.
\end{eqnarray}
\end{Theorem}
The projection of the initial data onto the  $(z_2,w_0)$-plane is a curve parameterized by $a$ that cuts through the saddle point corresponding to the SM in system (\ref{z2finalsumeqn112}), (\ref{w0finalsumeqn112}), between the stable and unstable manifold, (the dotted line in Figure 1). This implies that a small under-density corresponding to $a<1$ will  evolve to the stable rest point 
$(z_2,w_0)=(-4/3,1/3)$, (c.f. Figure 1).

\subsection{The Numerics}\label{S2.6}

 In this section we present the results of our numerical simulations.  We simulate solutions of (\ref{z2finalsumeqn112})-(\ref{w2finalsumeqn112}) for each value of the acceleration parameter $a<1$ in a small neighborhood of $a=1$, (corresponding to small under-densities  relative to the SM), and for each temperature $T_*$ in the range $3000^oK\leq T_*\leq 9000^oK$.  We simulate up to the time $t_a$, the time depending on the acceleration parameter $a$ at which the Hubble constant is equal to its present measured value $H=H_0=100h_0\frac{km}{s\,mpc}$, with $h_0=.68$.   From this we conclude that the dependence on $T_*$ is negligible.   We then asked for the value of $a$ that gives $Q(z_2(t_a),w_0(t_a))=.425$, the value of $Q$ in Dark Energy theory with $\Omega_{\Lambda}=.7$.  This determines the unique value $a=\underbar{\it a}=0.99999959$, and the unique time $t_0=t_{\underline{\it a}}$. These results are recorded in the following theorem:
 \begin{Theorem}  At present time $t_0$ along the solution trajectory of (\ref{z2finalsumeqn112})-(\ref{w2finalsumeqn112}) corresponding to $a=\underline{\it a}$, our numerical simulations give $H=H_0$, $Q=.425$, together with the following:
\begin{eqnarray}\nonumber
z(t_0,\xi)=(0.192)\xi^2+(2.871)\xi^4+O(\xi^6),
\end{eqnarray}
\begin{eqnarray}\nonumber
w(t_0,\xi)=(0.914)-(0.126)\xi^2+O(\xi^4),\ \ 
\end{eqnarray} 
and
\begin{eqnarray}\label{metricA}
A(t_0,\xi)=1-(0.064)\xi^2-(0.574)\xi^4,
\end{eqnarray} 
\begin{eqnarray}\label{metricD}
D(t_0,\xi)=1-(0.016)\xi^2+O(\xi^4).\ \ \ \ \  
\end{eqnarray} 
The cubic correction to redshift vs luminosity as predicted by the wave model at $a=\underline{\it a}$ is 
\begin{eqnarray}\label{valueofC}
C=0.359.
\end{eqnarray}
\end{Theorem}
Note that (\ref{metricA}) and (\ref{metricD}) imply that the spacetime is very close to Minkowski at present time up to errors $O(\xi^4)$, so the trajectory in the $(z_2,w_0)$-plane is much closer to the stable rest point $M$ than to the SM at present time, c.f. Figure 1. The cubic correction associated with Dark Energy theory with $k=0$ and $\Omega_{\Lambda}=.7$ is $C=-0.180$, so (\ref{valueofC}) is a theoretically verifiable prediction which distinguishes the wave theory from Dark Energy theory.   A precise value for the actual cubic correction corresponding to $C$ in the relation between redshift vs luminosity for the galaxies appears to be beyond current observational data.

\subsection{The Uniform Spacetime at the Center}\label{S2.7}

In this section we describe more precisely the central region of accelerated uniform expansion triggered by the instability due to perturbations that meet the ansatz (\ref{ansatz1})-(\ref{ansatz4}).  By (\ref{uniformdensity}) we have seen that neglecting terms of order $\xi^4$, the density $\rho(t)$ depends only on the time.   Further neglecting the small errors between $(z_2,w_0)$ and the stable rest point $\left(-\frac{4}{3},\frac{1}{3}\right)$ at present time $t_0$ when $a=\underbar{\it a}$, we prove that the spacetime is Minkowski with a density $\rho(t)$ that drops like $O(t^{-3})$, so the instability creates a central region that appears to be a flat version of a uniform Friedmann universe with a larger Hubble constant, in which the density drops at a faster rate than the $O(t^{-2})$ rate of the SM.  

Specifically, as $t\to\infty$, our orbit converges to $\left(-\frac{4}{3},\frac{1}{3}\right)$, the stable rest point for the $(z_2,w_0)$ system
\begin{eqnarray}\label{2levelode}
\left(\begin{array}{c}z_2\\w_0\end{array}\right)'
&=&\left(\begin{array}{c}-3w_0\left(\frac{4}{3}+z_2\right)\\-\frac{1}{6}z_2-\frac{1}{3}w_0-w_0^2\end{array}\right).
\end{eqnarray} 
Setting $z_2=-4/3+\bar{z}(t)$, $w_0=1/3+\bar{w}(t)$ and discarding higher order terms, we obtain the linearized system at rest point $(-\frac{4}{3},\frac{1}{3})$,
\begin{eqnarray}\label{restlin}
\left(\begin{array}{c}\bar{z}\\\bar{w}\end{array}\right)'
&=&\left(\begin{array}{cc}-1&0\\-\frac{1}{6}&-1\end{array}\right)\left(\begin{array}{c}\bar{z}\\\bar{w}\end{array}\right).
\end{eqnarray} 
The matrix in (\ref{restlin}) has the single eigenvalue $\lambda=-1$ with single eigenvector $R=(0,1)$.   From this we conclude that all orbits come into the rest point $(-\frac{4}{3},\frac{1}{3})$ from below along the vertical line $z_2=-4/3$.   This means that $z_2(t)$ and $\rho(t)=z_2(t)/t^2$ can tend to zero at algebraic rates as the orbit enters the rest point, but $w_0(t)$ must come into the rest point exponentially slowly, at rate $O(e^{-t})$.   Thus our argument that $\bar{w}=w_0-1/3$ is constant on the scale where $\rho(t)=k_0/t^{\alpha}$ gives the precise decay rate,
\begin{eqnarray}\label{restpointdensityfinal}
\rho(t)=\frac{k_0}{t^{3(1+\bar{w})}}.
\end{eqnarray}
That is, $\bar{w}\equiv\bar{w}(t)\to0$ and $k_0\equiv k_0(t)$ are changing exponentially slowly, but the density is dropping at an inverse cube rate, $O(1/t^{3(1+\bar{w})})$, which is {\it faster} than the $O(1/t^2)$ rate of the standard model. 

Therefore, neglecting terms of order $\xi^4$ together with the small errors between the metric at present time $t_0$ and the stable rest point, the spacetime is Minkowski with a density $\rho(t)$ that drops like $O(t^{-3})$, a faster rate than the $O(t^{-2})$ of the SM.    Furthermore, we show that neglecting relativistic corrections to the velocity of the fluid near the center where the velocity is zero,  evolution toward the stable rest point creates a flat, center independent spacetime which evolves outward from the origin, and whose size is proportional to the Hubble Length.  

We conclude that the effect of the instability triggered by a perturbation of the SM consistent with ansatz (\ref{ansatz1})-(\ref{ansatz4}) near the stable rest point $\left(-\frac{4}{3},\frac{1}{3}\right)$, is to create an anomalous acceleration consistent with the anomalous acceleration of the galaxies in a large, flat, uniform, center-independent spacetime, expanding outward from the center of the perturbation.

\section{Conclusion}\label{S4}  
The mechanism introduced here for the creation of the anomalous acceleration is derived from a rigorous self-contained mathematical model which identifies an unstable mode in the SM on the length scale of the supernova data. The resolution of this instability creates the same anomalous accelerations as the cosmological constant, without assuming it.   The model makes testable predictions.  If correct, it would imply that we live within a large region of approximate uniform under-density that is expanding outward from us at an accelerated rate relative to the SM.   The idea that the Milky Way lies near the center of a large region of under-density has already been proposed and studied in the physics literature. (See \cite{cliffela} and references therein.)     

The central region created by the instability\footnote{Given the instability of the Friedmann spacetime with respect to small perturbations, one has to wonder about the validity of the assumption that the universe is a Friedmann spacetime even on the scale of the supernova date, with or without Dark Energy.} is different from, but looks a lot like, a speeded up Friedmann universe tending more rapidly to flat Minkowski space than the SM.  The model is based on the starting assumption that Einstein's equations are correct without the cosmological constant. The result is a verifiable mathematical explanation for the anomalous acceleration of the galaxies that does not invoke Dark Energy. 

At this stage we have made no assumptions regarding the space-time far from the center of the perturbations that trigger the instabilities in the SM.  We have addressed one issue, the anomalous acceleration.  The consistency of this model with other observations in astrophysics would require additional assumptions.\footnote{No verifiable model in cosmology currently accounts for all of the physics,  \cite{stei}.  We need only point out the as yet unaccounted for large scale aspherical anomalies observed in the microwave background radiation, \cite{copihudjgd}.  We refer the interested reader to \cite{cliffela} for recent attempts to reconcile other under-density theories, with the standard $\Lambda CDM$ model of cosmology.}

\vspace{3cm}

 \begin{figure}
\centering
\caption{Phase Portrait for Central Region}
\includegraphics[width=1\textwidth]{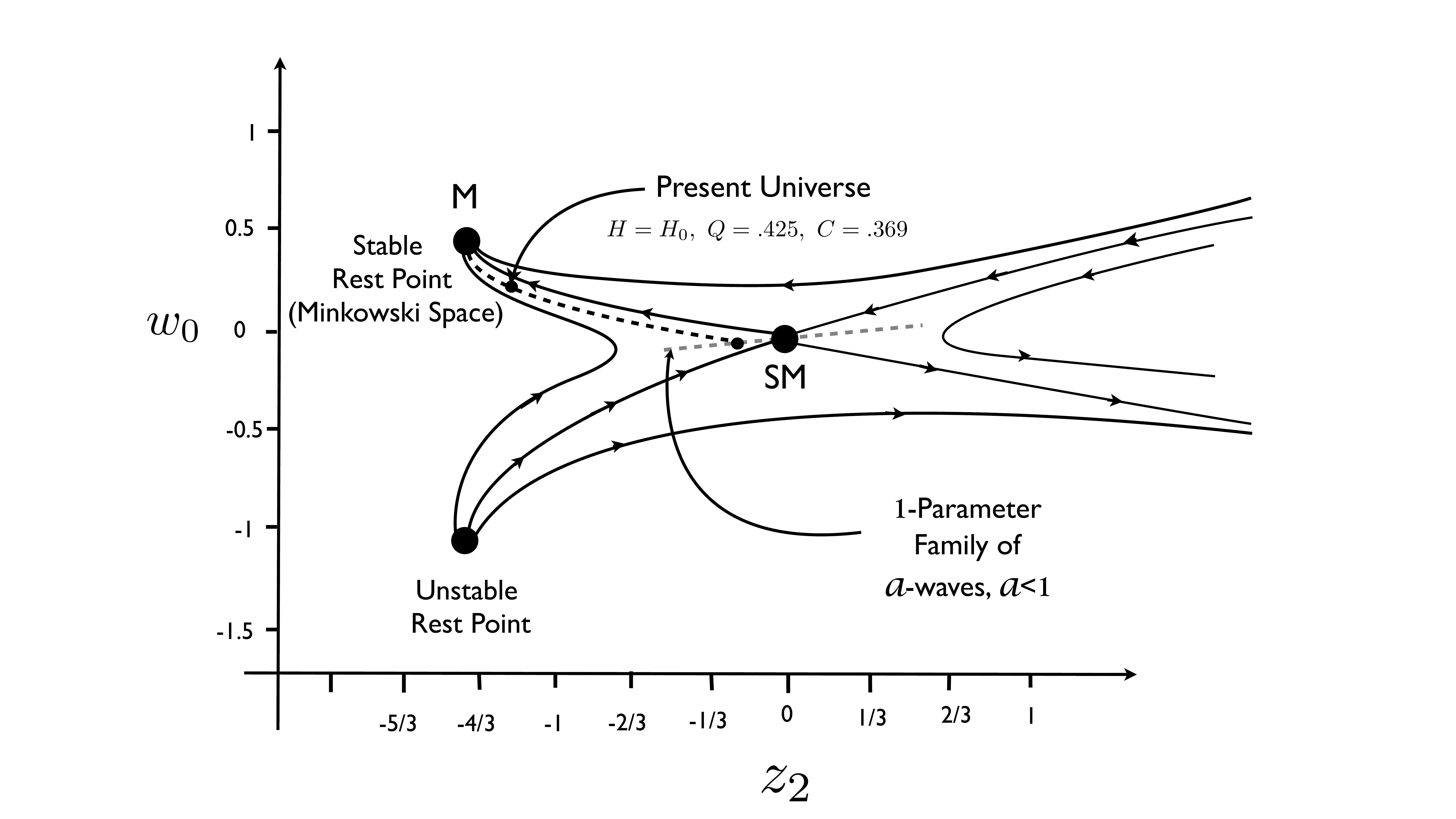}
\end{figure}

\end{document}